\documentclass[aps,prb,superscriptaddress,showpacs,twocolumn,a4paper]{revtex4-2}
\usepackage{babel}
\usepackage{amsmath,amsthm,amssymb,bm}
\usepackage{float}
\usepackage[pdftex]{graphicx}
\usepackage[utf8]{inputenc}
\usepackage{xcolor}
\usepackage{mathtools}

\providecommand{\includegraphics}[2][width=\textwidth]{$#2$}

\definecolor{citecol}{rgb}{0.152, 0.574, 0.664}
\definecolor{urlcol}{rgb}{0.152, 0.574, 0.664}
\usepackage[colorlinks=true, urlcolor=urlcol,citecolor=citecol]{hyperref}
\newcommand{\ua}{\uparrow}
\newcommand{\da}{\downarrow}

\begin{document}
\title{Order-by-disorder from Schwinger bosons in a frustrated honeycomb ferromagnet}
\author{Arnaud Ralko}
\affiliation{Institut N\'eel, Université Grenoble Alpes and CNRS, Grenoble, France}
\author{Jaime Merino}
\affiliation{Departamento de F\'isica Te\'orica de la Materia Condensada, Condensed Matter Physics Center (IFIMAC) and Instituto Nicol\'as Cabrera, Universidad Aut\'onoma de Madrid, Madrid 28049, Spain}

\date{\today}
\begin{abstract}
The cobalt-based honeycomb magnet BaCo$_2$(AsO$_4$)$_2$ (BCAO) has recently emerged as a promising platform for frustrated magnetism beyond conventional paradigms. 
Neutron-scattering experiments and first-principles calculations have revealed an unexpected double-zigzag (dZZ) magnetically ordered ground state, whose microscopic origin remains under active debate. 
Here, we revisit this problem within a ferro--antiferromagnetic $J_1$--$J_3$ Heisenberg model on the honeycomb lattice using a generalized Schwinger-boson mean-field theory (gSBMFT) that treats ferromagnetic and antiferromagnetic interactions on equal footing. 
This approach, combined with exact diagonalization (ED), allows us to demonstrate the emergence of the dZZ phase in a narrow parameter range, stabilized by quantum fluctuations through an order-by-disorder mechanism, in good agreement with recent density-matrix renormalization-group (DMRG) results. 
We further characterize the associated magnetic excitations and discuss their relevance to recent inelastic neutron-scattering (INS) measurements on BCAO.
\end{abstract}

\maketitle

\section{Introduction}

The investigation of frustrated magnetism in honeycomb materials has been fueled over the past decade by the search for the Kitaev spin liquid (KSL). 
The KSL is characterized by exotic excitations such as $\mathbb{Z}_2$ gauge fluxes and itinerant Majorana fermions, as well as an anomalous thermal Hall effect under an external magnetic field. 
Although the Ru-based material $\alpha$-RuCl$_3$ was initially considered an ideal platform for Kitaev physics, additional interactions drive the system toward more conventional magnetically ordered phases \cite{chaloupka_kitaev-heisenberg_2010,sinn_electronic_2016,yokoi_half-integer_2021,kasahara_majorana_2018}. 
This has motivated the search for alternative Kitaev materials such as cobaltates \cite{songvilay_kitaev_2020}, which, despite hosting weaker spin--orbit coupling (SOC) than their Ru (4$d$) and Ir (5$d$) counterparts, can realize Kitaev bond-dependent interactions \cite{maksimov_baco_2aso_4_2_2025,devillez2025bonddependentinteractionsillorderedstate}. 
In this regard, BaCo$_2$(AsO$_4$)$_2$ (BCAO) is possibly the most promising Co-based honeycomb magnet to host a KSL. 
Elastic neutron scattering on this 3$d^7$ KSL candidate indicates that it is an incommensurate antiferromagnet (AFM) \cite{regnault_book_1990} for $T<T_N=5.35$~K with ordering wavevector ${\bf Q}=(0.27,0,-1.31)$. 
For $T<T_N$, applying an external magnetic field $\mu_0H \approx \mu_0H_{c1}=0.33$~T along the in-plane $b$ direction of the honeycomb lattice drives BCAO into a commensurate magnetic phase with wavevector ${\bf Q}=(1/3,0,-1.31)$, which becomes almost fully polarized at $\mu_0H_{c2}=0.55$~T. 
On the other hand, inelastic neutron-scattering (INS) experiments on BCAO at $T=1.5$~K reveal a characteristic gapped flat mode \cite{devillez2025bonddependentinteractionsillorderedstate} with a minimum at the $\Gamma$ point, instead of the expected Goldstone mode at ${\bf Q}$. 
Together with the extremely small values of $T_N$ and $H_{c2}$, this observation suggests that BCAO lies close to a ferromagnetic (FM) transition in two dimensions \cite{maksimov_baco_2aso_4_2_2025}.

Despite the considerable amount of work devoted to BCAO and cobaltates in general, constructing a minimal model which captures all their fascinating magnetic properties has remained elusive. 
Previous studies combining {\it ab initio} calculations \cite{maksimov_ab_2022} with experimental observations \cite{maksimov_proximity-induced_2023,liu_non-kitaev_2023,lee_quantum_2025,das_xy_2021} support an XXZ-$J_1$-$J_3$ model, with FM $J_1<0$ and AFM $J_3>0$, supplemented by additional third-neighbor off-diagonal couplings $J^{(3)}_{z\pm}$ (expressed in the crystallographic frame), as a minimal description for BCAO. 
The ground state of this model is an unconventional $uudd$, or double zigzag (dZZ), state when $x = J_3/|J_1| \sim 0.2$--$0.4$ and $J^{(3)}_{z\pm} \neq 0$. 
More recent refinements have highlighted the importance of Kitaev bond-dependent exchange interactions \cite{maksimov_baco_2aso_4_2_2025} in order to reproduce the full details observed in INS experiments on BCAO. 
Monte Carlo simulations of a $J_1$-$K$-$\Gamma$-$\Gamma'$-$J_2$-$J_3$ model reconcile the nearly perfect dZZ order with the experimentally observed incommensurate magnetism and the gap $\Delta \sim 1.45$~meV at the $\Gamma$ point of the $ab$ plane in INS spectra. 
These calculations further suggest that the absence of a Goldstone mode \cite{devillez2025bonddependentinteractionsillorderedstate} can be attributed to defective dZZ order.
A common feature of these minimal models is their ability to accommodate the dZZ phase or closely related variants. 
Since the dZZ order represents a compromise between ferromagnetic (FM) and zigzag (ZZ) states, it is extremely sensitive to small magnetic fields, undergoing a transition to an intermediate $uud$ state before becoming fully polarized. 
This mechanism may account for the experimentally observed $1/3$ magnetization plateau in the field range $H_{c1} < H < H_{c2}$.
 
Motivated by recent DMRG calculations of the ferro--antiferromagnetic $J_1$--$J_3$ Heisenberg model on the honeycomb lattice, which revealed an order-by-disorder mechanism underlying the stabilization of the dZZ phase~\cite{jiang_quantum_2023}, we explore whether such a ``simple'' model already captures the essential ingredients of the magnetic properties of BCAO. 
Inspired by an extended version of SBMFT that treats singlet and triplet operators on equal footing~\cite{niggemann_quantum_2023,ralko_chiral_2024,Taran2026}, we employ a generalized SBMFT (gSBMFT) together with exact-diagonalization (ED) techniques to demonstrate how a dZZ state emerges in this model. 
While SBMFT~\cite{auerbach_book_2012} has been predominantly applied to frustrated antiferromagnetic systems~\cite{arovas_functional_1988,sachdev_kagome_1992}, we extend it here to treat frustrated ferromagnetic interactions, thereby overcoming the difficulties associated with violations of the boson number constraint~\cite{feldner_ferromagnetic_2011}. 
This extended framework enlarges the variational space and enables the stabilization of competing magnetic states beyond those captured within standard SBMFT.

\begin{figure}[ht!] 
  \centering
  \includegraphics[width=0.45\textwidth]{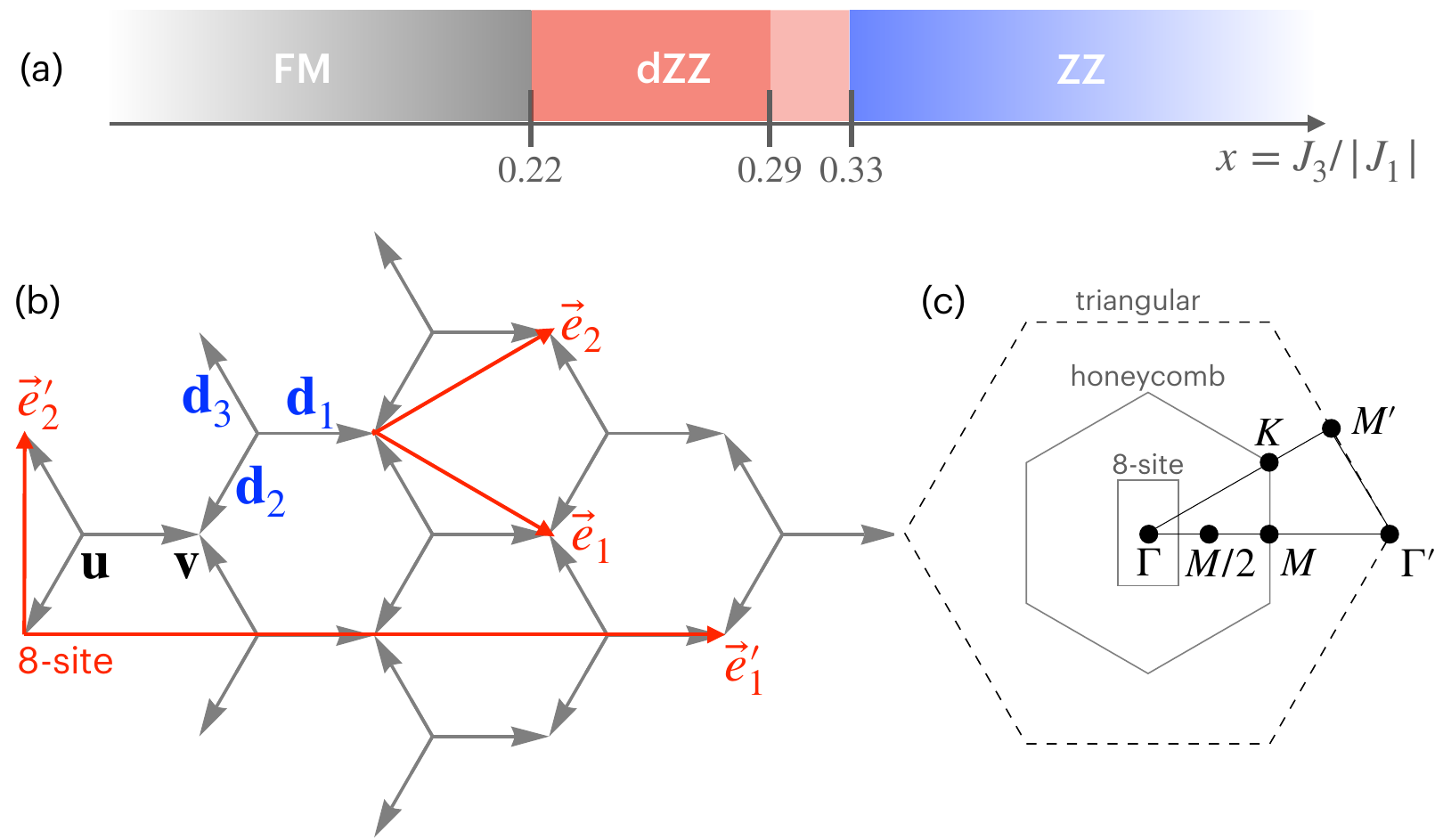}
  \caption{(a) Phase diagram of the ferro-antiferromagnetic $J_1$-$J_3$ model at $T=0$. (b) Honeycomb lattice with two sub-lattices $u$ and $v$, and the three nearest-neighbor displacement vectors $\{ \bf{d}_i \}$ for a 2-site unit-cell  and its primitive vectors $(\bf{e}_1,\bf{e}_2)$, and for an 8-site unit-cell with $(\bf{e}_1',\bf{e}_2')$. 
  (c) Corresponding first Brillouin zone of the honeycomb lattice (solid lines) and of the underlying triangular lattice defined by $(\vec{d}_1,\vec{d}_2)$ (dashed lines) together with high-symmetry points. The dynamical structure factors are analyzed along the path $\Gamma \frac{M}{2} M \Gamma' M' K \Gamma$, $\frac{M}{2}$ denoting the midpoint of the $\Gamma$–$M$ path.}
  \label{fig01}
\end{figure}
The phase diagram obtained within gSBMFT, shown in Fig.~\ref{fig01}(a), 
displays ferromagnetic (FM), zigzag (ZZ), and double-zigzag (dZZ) phases. 
While FM and ZZ orders are already present at the classical level, the stability and extent of the ZZ phase are significantly renormalized by quantum fluctuations. 
In particular, the dZZ phase emerges in an intermediate region $x \sim 0.22(5)$--$0.29(5)/0.33(1)$, where the two upper bounds correspond to its crossing with the ZZ phase obtained within standard SBMFT (denoted s-ZZ) and within gSBMFT (denoted g-ZZ), respectively.  
Our results broadly support the idea that the dZZ phase in the $J_1$--$J_3$ model is stabilized by quantum fluctuations through an order-by-disorder mechanism proposed recently \cite{jiang_quantum_2023} in the following sense. Classically, a spiral phase is the ground state of the model between the FM and ZZ phases. The latter phases become degenerate with the dZZ phase at $x=1/3$ with an energy which is just above the classical spiral energy. Our work indicates that the zero-point quantum corrections are dominant in the dZZ phase overcoming the classical
energy difference with the spiral ground state. This leads to the stabilization of the dZZ phase against the spiral or ZZ phases. 
Although the dZZ is selected from a classical unstable manifold it
is a result of an order-by-disorder mechanism leading to a collinear state 
which is unreachable at the classical level.  

The corresponding self-consistent solutions also reveal the nontrivial structure
of this peculiar magnetic order within the Schwinger-boson framework.

The remainder of this paper is organized as follows. 
In Sec.~\ref{sec:model}, we introduce the ferro--antiferromagnetic $J_1$--$J_3$ Heisenberg model on the honeycomb lattice and describe in detail the generalized Schwinger-boson mean-field theory (gSBMFT) developed to treat competing FM and AFM couplings on equal footing. 
In Sec.~\ref{sec:mean-field}, we present the resulting mean-field spinon structures and discuss the characteristic Ans\"atze corresponding to the ZZ and dZZ magnetic orders, emphasizing the role of triplet pairing terms in stabilizing the latter. 
Section~\ref{sec:phased} establishes the ground-state phase diagram by comparing the gSBMFT results with those of conventional SBMFT and exact-diagonalization calculations. 
This section also highlights the emergence of the dZZ phase through a quantum order-by-disorder mechanism \cite{jiang_quantum_2023} and analyzes the associated real-space spin-correlation patterns. 
In Sec.~\ref{sec:sq}, we compute the dynamical magnetic excitation spectra in both the ZZ and dZZ phases, identifying characteristic redistributions of spectral weight across Brillouin zones and discussing their relevance to recent inelastic neutron-scattering experiments on BaCo$_2$(AsO$_4$)$_2$. 
Finally, Sec.~\ref{sec:conclusion} summarizes our main conclusions and outlines possible extensions of the present work, including the effects of external magnetic fields, bond-dependent exchange interactions, and structural disorder on the stability of the dZZ phase.

\section{Model and methods}
\label{sec:model}

We consider the honeycomb lattice where, as shown in Fig.~\ref{fig01}(b), each lattice site of the $u$ sublattice is connected to three nearest neighbors on the $v$ sublattice through the bond vectors ${\bf d}_{1,2,3}$. 
We adopt an eight-site unit cell defined in Fig.~\ref{fig01}(b) and whose corresponding first Brillouin zone is also displayed in panel (c), together with the high-symmetry points encountered along the path used for the calculation of the dynamical properties. 
Note that $M/2$ denotes the midpoint of the $\Gamma$–$M$ path and corresponds to the condensation wavevector of the dZZ order.

We consider an isotropic Heisenberg model with interactions extending beyond nearest neighbors. Quantum spins ${\bf S}_i$ are located at each lattice site $i$, and the Hamiltonian reads
\begin{equation}
\label{eq:model}
{\cal H} = J_1 \sum_{\langle i,j \rangle} {\hat {\bf S}}_i \cdot {\hat {\bf S}}_j 
        + J_3 \sum_{\langle\langle\langle i,j \rangle\rangle\rangle} {\hat {\bf S}}_i \cdot {\hat {\bf S}}_j,
\end{equation}
where the first and second sums run over nearest- and third-neighbor pairs, respectively. 
Both $J_1$ and $J_3$ preserve full spin-rotation symmetry, in contrast to
bond-dependent couplings, which reduce lattice symmetries
\cite{maksimov_ab_2022,maksimov_baco_2aso_4_2_2025,maksimov_proximity-induced_2023}.
Frustration arises from the competition between ferromagnetic
nearest-neighbor interactions, $J_1<0$, and antiferromagnetic third-neighbor
interactions, $J_3>0$, as controlled by the ratio $x=J_3/|J_1|$.
The  competition between FM, $J_1<0$, and AFM, $J_3>0$, leads to frustration
to frustration in a $x = J_3 / |J_1|$ parameter range which is explored here.  
We analyze Eq.~(\ref{eq:model}) within the Schwinger-boson mean-field theory (SBMFT) framework~\cite{auerbach_book_2012,arovas_functional_1988}, 
a method that has been extensively applied to frustrated AF Heisenberg models~\cite{sachdev_kagome_1992,lugan_schwinger_2022,halimeh_spin_2016,kos_quantum_2017,schneider_projective_2022,kargarian_unusual_2012}. 
Prominent examples include the AF $J_1$--$J_2$ model on the square lattice~\cite{fa_wang_square_2016}, as well as the triangular, kagome~\cite{sachdev_kagome_1992,fa_wang_kagome_2006}, and honeycomb lattices~\cite{fa_wang_honeycomb_2010,merino_role_2018}. 
In contrast, systems involving both FM and AFM couplings have been explored much less within SBMFT on finite clusters due to the violation of the boson number constraint  \cite{feldner_ferromagnetic_2011} even on average. Our model, Eq.~(\ref{eq:model}), explicitly includes competing FM and AFM interactions on the honeycomb lattice, thereby extending the applicability of SBMFT to a regime rarely addressed in previous studies, apart from, {\it e.g.}, related work on the square lattice~\cite{sarker_bosonic_1989,feldner_ferromagnetic_2011}, ferrimagnetism in one-dimensional chains~\cite{wu_schwinger-boson_1999}, or more advanced theoretical developments~\cite{zhang_schwinger_2022,yan_combined_2015,trumper_schwinger-boson_1997,tay_variational_2011}.

In the Schwinger boson representation, the spin operator at site $i$ is expressed as
\begin{eqnarray}
\hat{\bf S}_i &=& \frac{1}{2} \hat b_{i}^\dagger {\boldsymbol \sigma} \hat b_{i},
\end{eqnarray}
where $\hat b_i^\dagger = (\hat b_{i,\uparrow}^\dagger,\hat b_{i,\downarrow}^\dagger)$, the boson spinor,  creates bosonic spinons, and ${\boldsymbol \sigma} = \sigma^1 {\bf u}_x + \sigma^2 {\bf u}_y + \sigma^3 {\bf u}_z$ denotes the Pauli matrices. We also define $\sigma^0$ as the $2\times 2$ identity matrix, with $\alpha=0,1,2,3$ labeling the components.

To reproduce the correct spin-$S$ algebra, one must impose the boson-number constraint
\begin{eqnarray}
\label{eq:constraint}
\hat{n}_i = \hat b_{i}^\dagger \hat b_{i} = 2 S,
\end{eqnarray}
which projects out unphysical states. While exact enforcement would render the representation faithful, in practice it is too demanding \cite{wang_schwinger-boson_2000}, and the constraint is implemented on average by introducing Lagrange multipliers:
\begin{eqnarray}
\sum_i \lambda_i \left( \hat{n}_{i } - 2S \right),
\end{eqnarray}
added to the Hamiltonian.
In practice, translational invariance is imposed at the level of the chosen unit cell, so that the Lagrange multipliers are taken to be identical on equivalent sites, $\lambda_i \equiv \lambda_a$, where $a$ labels the inequivalent sites of the unit cell. This results in a set of independent parameters $\{\lambda_a\}$, rather than a single global $\lambda$.
Within this framework, $S$ can be viewed as a parameter controlling the strength of quantum fluctuations, allowing us to explore the quantum melting of magnetically ordered states and their connection to quantum spin-liquid phases.

The Hamiltonian (\ref{eq:model}) contains only SU(2)-invariant Heisenberg terms that take the generic form,
\begin{eqnarray}
\label{identities}
\hat {\bf S}_i \hat{J} \hat{\bf S}_j &=& \frac{1}{4} J_{\alpha,\beta} 
   T^{\alpha \beta}_{s_0 s_1 s_2 s_3} 
   \hat b_{i,s_0}^\dagger \hat b_{i,s_1} \hat b_{j,s_2}^\dagger \hat b_{j,s_3},
\end{eqnarray}
where repeated indices are summed up, $J$ is the $3 \times 3$ coupling matrix on the bond $(i,j)$, and $T^{\alpha \beta}_{s_0 s_1 s_2 s_3} = \sigma^\alpha_{s_0,s_1} \sigma^\beta_{s_2,s_3}$ is a rank-4 tensor. Up to this stage, the parton construction remains exact, independent of the form of $J$.

To proceed further, we approximate the quartic operators by a
multichannel mean-field decoupling in both pairing and hopping bond
channels. Schematically, this Hartree--Fock--Bogoliubov decoupling reads
\begin{eqnarray}
 \hat b_{i,s_0}^\dagger \hat b_{i,s_1}
 \hat b_{j,s_2}^\dagger \hat b_{j,s_3}
 &\simeq&
 ( \hat b_{i,s_0}^\dagger \hat b_{j,s_2}^\dagger
 \, | \,
 \hat b_{i,s_1} \hat b_{j,s_3} ) \nonumber  \\
 &+&
 ( \hat b_{i,s_1} \hat b_{j,s_2}^\dagger
 \, | \,
 \hat b_{i,s_0}^\dagger \hat b_{j,s_3} ) .
\label{eq:general_decoupling}
\end{eqnarray}
Here
\begin{equation}
(\hat X | \hat Y)
=
\langle \hat X \rangle \hat Y
+
\hat X \langle \hat Y \rangle
-
\langle \hat X \rangle \langle \hat Y \rangle
\end{equation}
denotes the linearization of a product of two bilinear operators around
their saddle-point expectation values. Equation~(\ref{eq:general_decoupling})
should therefore be understood as a mean-field approximation, not as an
operator identity.

The pairing and hopping terms in Eq.~(\ref{eq:general_decoupling}) are
not independent copies of the same interaction. They correspond to
distinct particle-particle and particle-hole channels generated by the
original spin-exchange tensor. Their relative weights are fixed by the
microscopic interaction and by the channel decomposition, rather than
introduced as additional phenomenological parameters. The corresponding
pairing $\langle \hat b_i \hat b_j\rangle$ and hopping
$\langle \hat b_i^\dagger \hat b_j\rangle$ mean-field amplitudes are then
determined self-consistently in the Bogoliubov boson vacuum.

Additional symmetry constraints can be imposed on the mean-field vectors,
reducing the generalized decoupling to more restricted SBMFT forms
(see App.~A for an example).

Our gSBMFT formulation is more general than the SBMFT since it does not anticipate any specific forms for the operators, and all possible spin combinations can be active in the theory. Note that enlarging the number of mean-field parameters beyond the SBMFT does not mean that we are introducing free parameters in an uncontrolled way. The gSBMFT introduced here is based on a well justified mean-field decoupling allowing to access broken symmetry solutions not included in the SBMFT. In particular, one could deal with singlet or triplet pairing/hopping independently, and consider any kind of spin couplings on a given bond. Note that while being equivalent with  \cite{ralko_chiral_2024}, the present formalism is more compact in notation. Also, it can be useful to make contact with the physical hopping and pairing operators defined as 
\begin{eqnarray}
\hat{h}_{ij}^\alpha = \frac{1}{2}  {\hat b}_{i}^+  \sigma^{\alpha} {\hat b}_{j} ,~~~~\hat{p}_{ij}^\alpha = \frac{i}{2}  {\hat b}_{i} \left( \sigma^\alpha  \sigma^2 \right) {\hat b}_{j },
\end{eqnarray}  with $\alpha = 0$ belonging to the singlet  and $\alpha = 1,2,3$ triplet channels.
 This includes recent works where more than two operators (singlet pairing and spinon hopping) were considered, necessary when dealing with {\it e.g.} Dzyaloshinskii-Moriya \cite{manuel_heisenberg_1996,messio_schwinger-boson_2010,messio_chiral_2017,mondal_schwinger_2017} of anisotropic interactions like in Kitaev systems \cite{kos_quantum_2017,ralko_chiral_2024,sasamoto2025schwingerbosontheorys1}. More details on the theory are provided in the appendices.

Solving the resulting mean-field hamiltonian self-consistently, a saddle point is reached at which the corresponding Ansatz should satisfy the relation $\langle (\hat{X} | \hat{Y} ) \rangle  =  \langle \hat{X} \rangle \langle \hat{Y} \rangle$. Hence, a valid Ansatz, should simultaneously satisfy this relation as well as the boson constraint Eq.~(\ref{eq:constraint}).

A key aspect of the present model is that it contains both FM $J_1$, and AFM $J_3$ interactions. 
In the SBMFT treatment of the $T=0$ FM Heisenberg model, all bosons condense at the $\Gamma$ point, consistent with the expected classical FM order. Our present SBMFT implementation on finite lattices analogous to previous works \cite{lamas_sbmft_2013} does not consider Bose condensation explicitly. In our approach, as the system is increased, a macroscopic boson fraction gradually builds up at the $\Gamma$ point which is gapped 
when the singlet paring amplitudes are nonzero signaling Bose condensation in the thermodynamic limit. However in the present SBMFT treatment of the FM case, the singlet pairing amplitudes vanish and only hopping amplitudes, {\it i.e.} the triplet $h_{ij}$ amplitudes, are nonzero. This leads to a 
SBMFT hamiltonian which conserves the boson particle number with, however, a vacuum violating the average boson number constraint. 
This is avoided in our extension,  gSBMFT~\cite{ralko_chiral_2024}, which includes both singlet and
triplet pairing amplitudes accounting for FM correlations in FM bonds. This allows for a description of FM correlations on finite lattices while imposing the average number of bosons per site. 
Hence, the gSBMFT has the potential to describe more complex magnetic states, since it involves combinations of singlet and triplet operators in both pairing and hopping channels~\cite{ralko_chiral_2024}. 
We follow this route here and compare both methods whenever possible.

\section{Mean-field spinon structures and dispersions}
\label{sec:mean-field}

Here we introduce the relevant spinon Ansätze used in the present study of model \eqref{eq:model} and analyze their mean-field structure as well as spinon dispersion relations.
\begin{figure}[ht!] 
  \centering
  \includegraphics[width=0.4\textwidth]{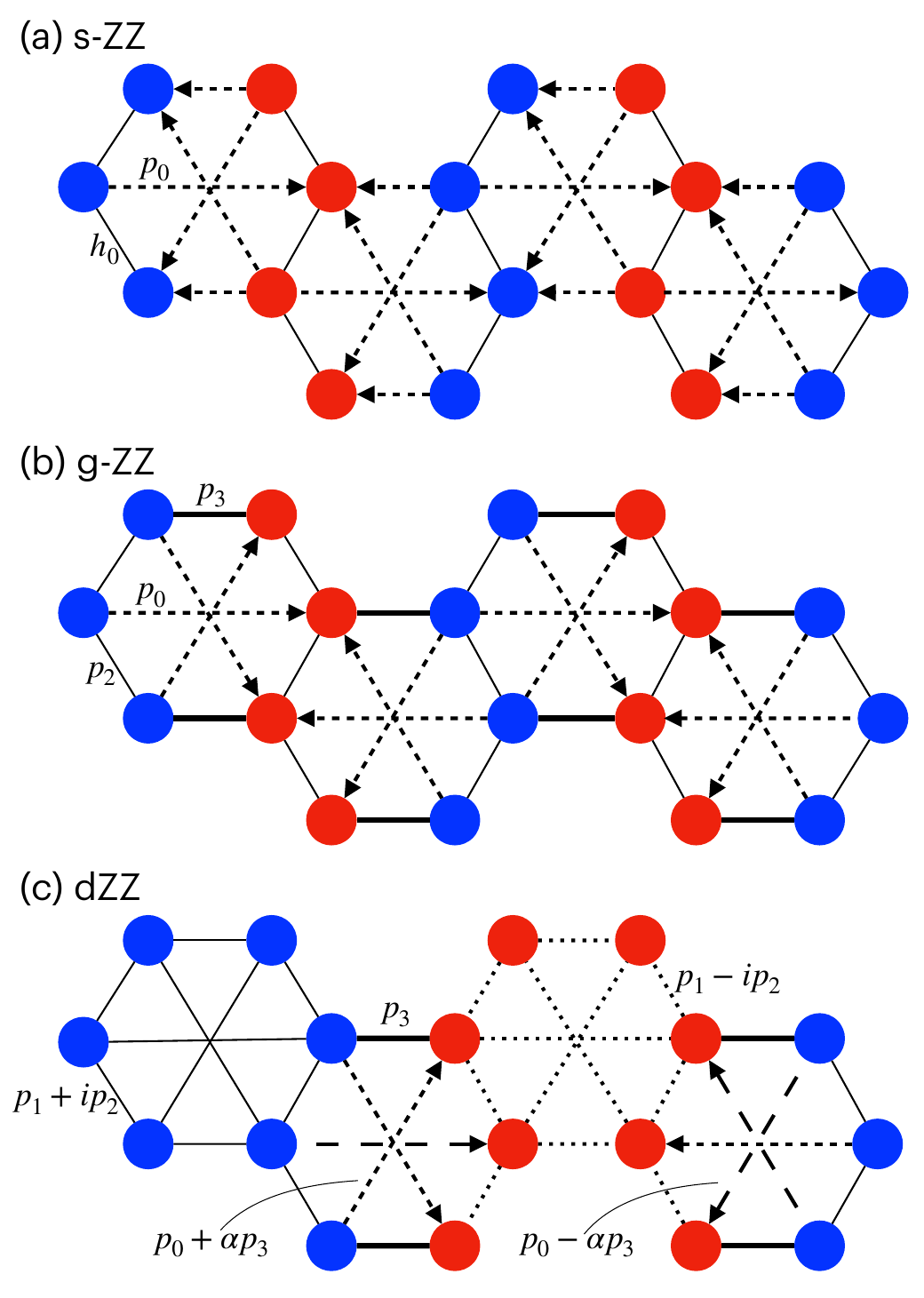}
\caption{Spinon Ans\"atze of the ZZ and dZZ phases within standard and generalized SBMFT.
(a) Mean-field structure of the ZZ phase obtained within standard SBMFT (s-ZZ). Despite ferromagnetic nearest-neighbor bonds, finite singlet pairing amplitudes are present.
(b) Corresponding ZZ Ansatz within gSBMFT (g-ZZ), where ferromagnetic bonds (solid lines) host finite triplet pairing amplitudes $p_{2,3}\neq0$, while antiferromagnetic $J_3$ bonds (dashed lines/arrows) carry singlet pairing $p_0$. This Ansatz lies slightly higher in energy than (a); see text for discussion.
(c) Mean-field structure of the dZZ phase, obtained only within gSBMFT on the 8-site unit cell. In the self-consistent solution, only pairing amplitudes remain finite.
In all panels, blue and red sites indicate the two opposite spin orientations of the corresponding collinear magnetic pattern.}
  \label{fig02}
\end{figure}

Fig.~\ref{fig02} displays the mean-field spinon sign patterns corresponding to the ZZ and dZZ Ans\"atze, as obtained within the (a) SBMFT and (b,c) gSBMFT frameworks.  
In the conventional SBMFT formulation, all FM and AFM bonds—are described in terms of singlet pairing $p_0$ and spinon hopping $h_0$.  
In contrast, the gSBMFT Ansatz exhibits a crucial distinction: on ferromagnetic bonds, the triplet pairing amplitudes $p_{2,3}$ acquire finite expectation values.  
This allows stable self-consistent ZZ and dZZ solutions to be obtained, with converged energies satisfying the boson number constraint on average.  

We find a sequence of phase transitions with increasing $J_3$, from a ferromagnetic state at low $J_3$ to the dZZ phase, and finally to the ZZ phase, in agreement with previous analyses \cite{jiang_quantum_2023,maksimov_ab_2022,maksimov_baco_2aso_4_2_2025,maksimov_proximity-induced_2023}. 
Since the intermediate dZZ phase does not correspond to any of the energetically favored classical orders, it is selected by quantum fluctuations.  
Hence, the gSBMFT predicts a small but finite range of $J_3$ where the mechanism of quantum order-by-disorder becomes operative, corroborating previous DMRG findings on smaller systems \cite{jiang_quantum_2023}.

\begin{figure}[ht!]
  \centering
  \includegraphics[width=0.4\textwidth]{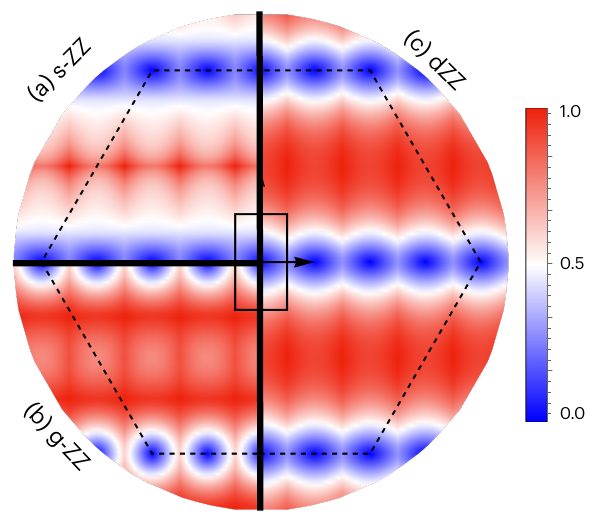}
\caption{
Lowest-energy spinon dispersions for (a) the s-ZZ phase obtained from SBMFT at $x=0.4$, (b) the g-ZZ phase obtained from gSBMFT at $x=0.4$, and (c) the dZZ phase obtained from gSBMFT at $x=0.25$. 
The calculations are performed at $S=1/2$ on the 8-site unit cell for a system size of $8\times12\times12$ sites. 
For all cases, only a portion of the Brillouin zone shown in Fig.~\ref{fig01} is displayed.
}
  \label{fig03}
\end{figure}

Magnetic order is inferred through finite-size scaling by tracking the emergence of Goldstone-mode precursors in the spinon dispersion. 
Within SBMFT, these modes generically appear at momenta $\pm {\bf Q}/2$, where ${\bf Q}$ is the ordering wavevector of the corresponding magnetic phase \cite{lamas_sbmft_2013}. 
In Fig.~\ref{fig03} we show the lowest-energy spinon dispersions obtained within the eight-site unit cell for $S=1/2$.
Panels (a) and (b) display the results for the s-ZZ and g-ZZ at $x=0.4$, while panel (c) shows the dispersion for the dZZ phase at $x=0.25$ within gSBMFT.
Although the g-ZZ solution is slightly higher in energy than s-ZZ, its dispersion is shown for comparison to highlight the effect of triplet channels on the spinon dispersion.
The minima of the lowest-energy spinon dispersion (blue regions) exhibit a characteristic stripe-like pattern corresponding to the ZZ and dZZ magnetically ordered vertical stripes. Such minima are located at the $\Gamma$ point of the reduced Brillouin zone (see Fig.~\ref{fig01}) of the 8-site unit cell. 

The lowest-energy spinon dispersions obtained within the SBMFT and gSBMFT Ans\"atze for the ZZ phase are shown in panels (a) and (b), respectively, while panel (c) displays the dispersion of the dZZ phase within gSBMFT.

Although the overall structure of the ZZ dispersions is similar in both approaches,
small differences can be observed. In contrast, the dZZ phase exhibits a clearly
distinct low-energy structure in the middle of the Brillouin zone. This feature
reflects the different sign patterns of the mean-field parameters encoded in the
two Ans\"atze (see Fig.~\ref{fig02}(b,c)).

\section{Phase diagram}
\label{sec:phased}

\begin{figure}[ht!] 
  \centering
  \includegraphics[width=0.45\textwidth]{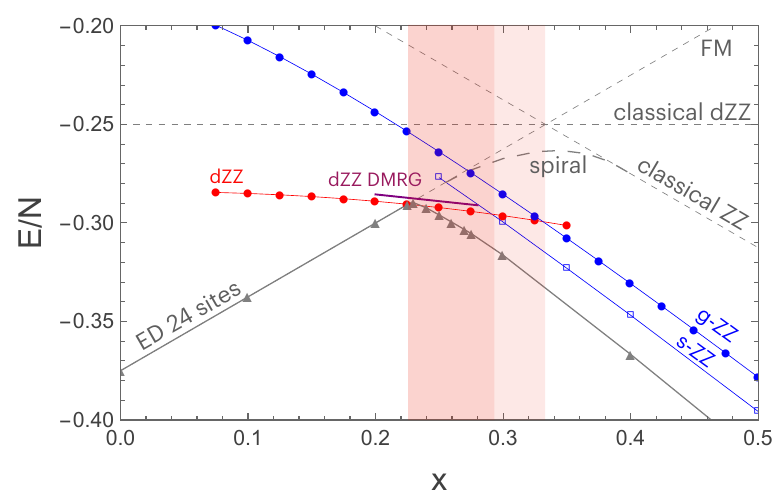}
\caption{Ground-state energy per site as a function of $x$ for $S=1/2$. 
Exact diagonalization (ED) results on a 24-site hexagonal cluster are shown (closed gray triangles) 
with the thick gray line a line for the eye. 
Mean-field results are shown for the g-ZZ (solid blue squares) and dZZ (solid red circles) Ans\"atze, 
together with the s-ZZ (open blue squares). 
DMRG data for the dZZ phase digitized from Fig.~3(b) of Ref.~\cite{jiang_quantum_2023} are shown in purple.  
Dashed gray lines indicate the classical energies of the FM, ZZ, and dZZ orders. The pink shaded regions indicate the finite parameter range where the dZZ phase is stabilized.
The darker region corresponds to its extent relative to the s-ZZ solution, while the lighter region shows the extension obtained when the ZZ phase is treated within the gSBMFT.}
  \label{fig04}
\end{figure}

We have obtained the phase diagram of model~(\ref{eq:model}) using both the standard and generalized SBMFT approaches on 8-site unit cells compatible with all Ans\"atze presented in the previous section, considering cluster size of $N = 8 \times 2 \times l^2$ sites, up to $l=12$. 
To support our results, we also perform ED calculations on 24-site hexagonal clusters with periodic boundary conditions (see Appendix \ref{app.d} for technical details), which confirm the emergence of the dZZ phase in an intermediate regime between the FM and ZZ phases.

A significant difference between the two mean-field approaches emerges when considering the dZZ phase.  
Within SBMFT, we have been unable to find a self-consistent dZZ solution. This may be attributed to the fact that, in the dZZ phase, each spin is FM coupled to all three nearest neighbors, similarly to the FM case. 
In contrast, the gSBMFT formulation introduced here allows for linear combinations of singlet and triplet components in both pairing and hopping channels, thereby significantly enlarging the variational space. 
This extension enables a self-consistent and energetically stable dZZ solution, occurring in a narrow parameter range between the FM and ZZ phases. 
These findings are consistent with recent DMRG studies performed on long cylinders~\cite{jiang_quantum_2023} as well as with our ED calculations discussed below.

In Fig.~\ref{fig04} we show the ground-state energy as a function of $x=J_3/|J_1|$ obtained from SBMFT, ED, and at the classical limit for $S=1/2$.
As discussed above, the SBMFT treatment on finite clusters fails to reproduce the pure FM state without violating the average boson-number constraint, as this state is purely classical. 
Similarly, the gSBMFT approach cannot adequately capture the FM phase despite its broader parameter space. In the pure FM case, we find that self-consistency drives all mean-field parameters in gSBMFT to zero even on small clusters. This reflects the classical nature of the FM state in which all bosons are condensed.
Therefore, we assume the exact classical energy dependence on $x$ in the FM phase. 
By comparing this energy with those of the ZZ obtained by both the standard and the generalized SBMFT, and dZZ phases, we can determine the phase boundaries of the model. 
Our ED calculations reveal an energy much lower compared to the classical energy for $x \gtrsim 0.225$, {\it i.e.} above the FM phase. This clearly indicates the prominent role played by quantum fluctuations. 
In particular, quantum fluctuations extend the parameter range in which the ZZ state is stable to lower $x$. 
Crucially, the dZZ state—rather than the classical spiral phase—emerges between the FM and ZZ phases. This can be interpreted from an order-by-disorder argument by which the zero-point quantum correction dominates in the dZZ phase compared to the ZZ, spiral or dZZ phases overcoming the energy difference with the spiral ground state.

The gSBMFT energies in Fig.~\ref{fig04} show that the dZZ phase becomes favorable below $x \lesssim 0.33(5)$ and $x \lesssim 0.29(5)$ with respect to the g-ZZ and s-ZZ solutions, respectively, while the FM state persists for $x \lesssim 0.225$. 
These values define the upper boundaries of the dZZ phase, set by its crossing with the two ZZ solutions, with g-ZZ lying slightly higher in energy than s-ZZ.
This defines two estimates for the upper boundary of the dZZ phase depending on the type of SBMFT used to describe the ZZ state. 
Overall, the gSBMFT energy dependence of the dZZ phase closely follows both the ED results near the FM transition and the recent DMRG data, digitized from Fig.~3(b) of Ref.~\cite{jiang_quantum_2023}  (purple in Fig.~\ref{fig04}). 
Note however that since SBMFT is intrinsically non-variational, the mean-field energies may occasionally lie slightly below those obtained from ED.

From our $T=0$ ground-state energy analysis we find that quantum fluctuations stabilize the dZZ order, which is not among the classical ground states of the model for any value of $x$. 
This indicates that the dZZ phase is selected through a quantum order-by-disorder mechanism as a compromise between the FM and ZZ phases. The zero-point quantum corrections are dominant in the dZZ phase overcoming the classical energy difference with the spiral ground state eventually stabilizing the dZZ state. This is because the zero point quantum fluctuations are larger in the collinear dZZ phase than the spiral phase. 

\begin{figure}[ht!] 
  \centering
  \includegraphics[width=0.45\textwidth]{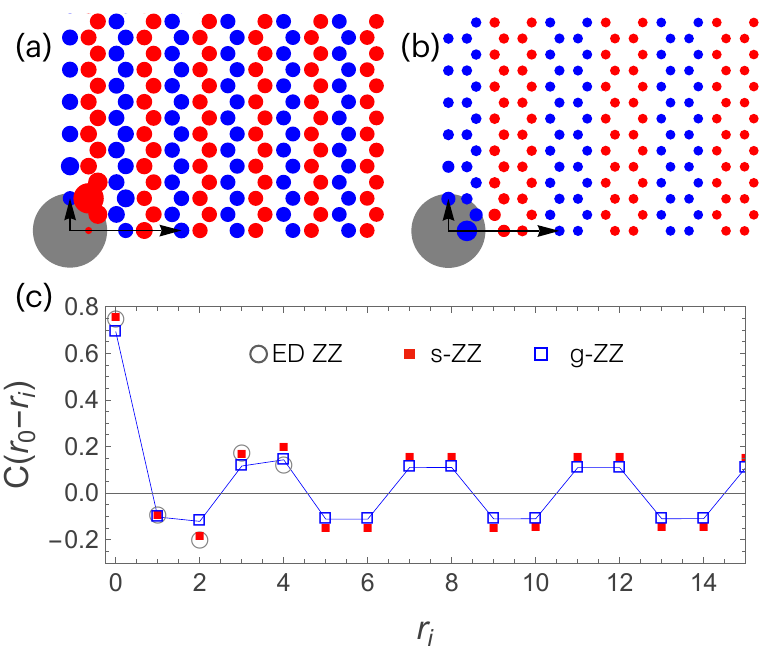}
\caption{Real-space spin--spin correlations at $S=\tfrac{1}{2}$. 
(a) ZZ phase at $x=0.4$ and (b) dZZ phase at $x=0.25$. 
Disk size indicates the correlation strength, while colors denote positive (blue) and negative (red) correlations. 
The reference site is shown in gray and black arrows indicate the translation vectors of the 8-site unit cell. 
(c) ZZ correlations obtained from ED on a 24-site ($8\times3$) cluster (gray circles), compared with SBMFT (red) and gSBMFT (blue) calculations performed on an $8\times2\times12\times12$ cluster at $x=1$.}
\label{fig05}
\end{figure}
Further insight into the role of quantum fluctuations in model~(\ref{eq:model}) can be obtained by analyzing the real-space spin correlations in the ground state. 
We define the spin correlations as
\begin{equation}
C(\mathbf{r}_i - \mathbf{r}_j) = \langle {\hat {\bf S}}_i \cdot {\hat {\bf S}}_j \rangle,
\label{eq:Cr}
\end{equation}
which characterize the spatial arrangement of the spins.

In Fig.~\ref{fig05}, we show the real-space correlations $C(\mathbf{r}_0 - \mathbf{r}_i)$ obtained for the g-ZZ [panel (a)] and dZZ [panel (b)] phases at $x=0.4$ and $x=0.25$, respectively, both chosen deep inside their stability regions. In particular, the dZZ phase is the ground state at $x=0.25$ within gSBMFT, as shown in Fig.~\ref{fig04}. The size of the circles represents the magnitude of the correlations with respect to a reference site located at the lower-left corner, while the color encodes their sign (blue for positive and red for negative correlations). It is immediately apparent that the ZZ phase exhibits a more robust magnetic order than the dZZ phase, which displays a weaker modulation of the correlations and a more uniform background.

Fig.~\ref{fig05}(c) shows the spin--spin correlation function of the ZZ phase along the armchair direction. The expected AFM alignment along the armchair direction, together with FM correlations along the zigzag chains, confirms the characteristic ZZ order. Details on the calculation of $C(\mathbf{r}_0 - \mathbf{r}_i)$ are provided in Appendix~\ref{app.c}. Within SBMFT, the standard and generalized approaches yield identical results, as illustrated in the figure. 

A direct comparison with ED results on the 24-site ($8\times3$) cluster  (see Appendix~\ref{app.d} for details) shows very good quantitative agreement, in particular in the decay and oscillatory behavior of the correlations with distance.

\begin{figure}[ht!] 
  \centering
  \includegraphics[width=0.45\textwidth]{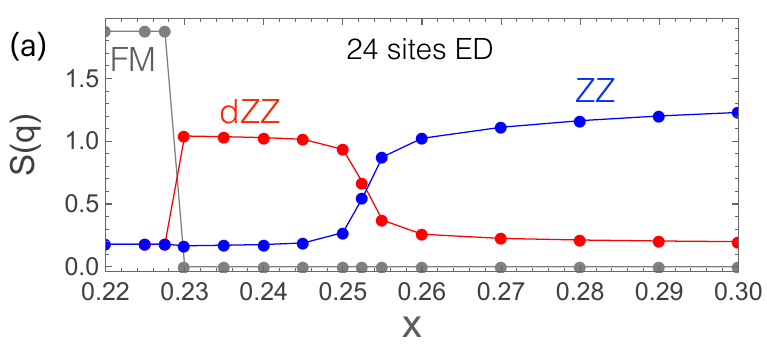}
\caption{Static spin structure factors from ED on a 24-site ($8\times3$) cluster at representative Brillouin-zone points: $\Gamma$ (gray), $M$ (blue), and $M/2$ (red), corresponding to the dominant Bragg peaks of the FM, ZZ, and dZZ phases.}
  \label{fig06}
\end{figure}

Fig.~\ref{fig06}(a) displays the static spin structure factor
\begin{equation}
S(\mathbf{q}) = \frac{1}{N} \sum_{i,j} e^{i \mathbf{q}\cdot(\mathbf{r}_i-\mathbf{r}_j)} 
\langle {\hat {\bf S}}_i \cdot {\hat {\bf S}}_j \rangle,
\label{eq:Sq}
\end{equation}
obtained from ED on the same 24-site cluster at representative $\mathbf{q}$ points in the Brillouin zone: $\Gamma$ (gray), $M$ (blue), and $M/2$ (red), corresponding respectively to the dominant Bragg peaks of the FM, ZZ, and dZZ phases. Even for this modest cluster size, the presence of the dZZ phase is clearly evidenced, and its extent as a function of $J_3$ is consistent with both the literature~\cite{jiang_quantum_2023} and our gSBMFT results.

Altogether, the gSBMFT framework introduced here is in agreement with ED and DMRG results. This provides further support to the order-by-disorder mechanism underlying the existence of the dZZ phase in the ferro-antiferromagnetic $J_1$-$J_3$ Heisenberg model since our results are obtained in much larger systems. Our work also shows how the gSBMFT is a better suited method than the SBMFT to deal with the magnetic properties of frustrated ferromagnets.

\section{Dynamical magnetic excitations}
\label{sec:sq}

As discussed above, the onset of magnetic order is signaled by the emergence of sharp features in the static structure factor $S(\mathbf{q})$ at the ordering wavevector ${\bf Q}$.
We now also analyze the associated magnetic excitation spectra in the different ordered phases based on our gSBMFT approach on finite clusters. For this purpose we analyze the dynamical structure factor:
\begin{equation}
S(\mathbf{q},\omega) = \frac{1}{2\pi N} \sum_{i,j} e^{i \mathbf{q}\cdot(\mathbf{r}_i-\mathbf{r}_j)} 
\int_{-\infty}^{\infty} dt \, e^{i\omega t} 
\langle {\hat {\bf S}}_i(t)\cdot {\hat {\bf S}}_j(0) \rangle.
\label{eq:Sqw}
\end{equation}
Magnetic order is signaled in our finite clusters by the presence of magnon-like excitations as the thermodynamic limit is approached.

\begin{widetext}
\begin{center}
\begin{figure}[!ht!]
  \includegraphics[clip,width=0.95\textwidth]{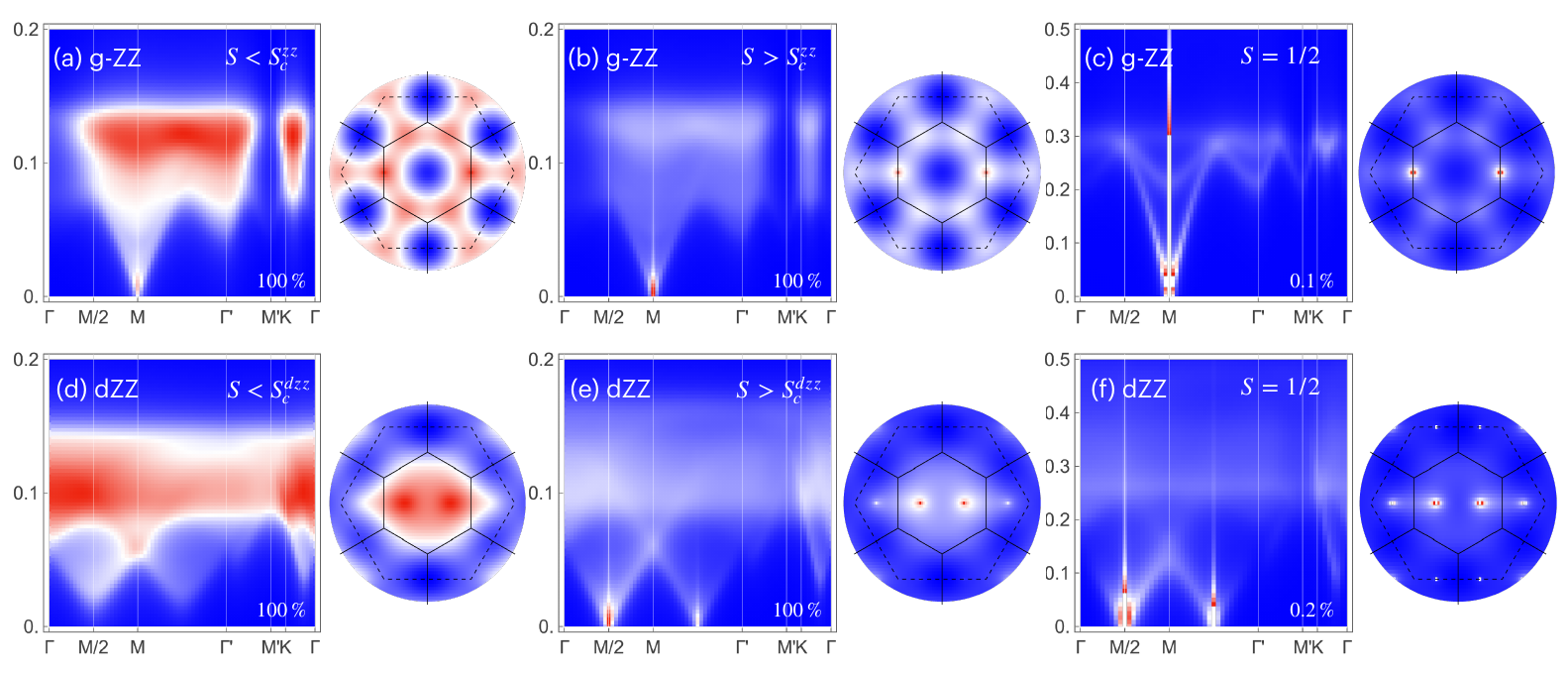}
\caption{
Normalized static and dynamical spin structure factors for the zigzag (a–c) and double-zigzag (d–f) phases. 
Panels (a,b) and (d,e) correspond to effective spin values slightly below and above the critical values $S_c^{zz}\simeq0.08$ and $S_c^{dzz}\simeq0.125$, respectively, but are computed without explicit boson condensation, so that the system remains in a non-condensed mean-field regime. 
Panels (c) and (f) show the physical case $S=1/2$. 
The zigzag phase is computed at $x=J_3/|J_1|=0.4$, while the double-zigzag phase is shown at $x=0.25$, on a system of $8\times2\times12\times12$ sites. 
For panels (a,b,d,e), the full spectral weight is displayed. In contrast, for the strongly ordered $S=1/2$ cases (c,f), only $0.1\%$ of the total spectral weight is shown to avoid saturation by the Bragg peak. 
This representation highlights the residual spectral weight arising from quantum fluctuations that persists even in these magnetically ordered states.
}
  \label{fig07}
\end{figure}
\end{center}    
\end{widetext} 

In Fig.~\ref{fig07}, we present the static and dynamical spin structure factors of the ZZ and dZZ phases obtained on an 8-site unit cell. 
All quantities are computed from the full expectation value of the bosonic operators in the Bogoliubov vacuum and evaluated for effective spin values $S$ in the vicinity of a critical value $S_c$, as well as for the physical case $S=1/2$. 
Within SBMFT, $S_c$ marks the closing of the spinon gap at the ordering wavevector and signals the onset of magnetic order, without explicitly introducing boson condensation. 
Magnetic order is thus inferred from the gap closing in this framework. 
Varying $S$ therefore provides a convenient way to probe the stability of competing phases. 
Technical details of the calculation are given in Appendix~\ref{app.c}.

More specifically, Fig.~\ref{fig07} compares the normalized static and dynamical structure factors of the g-ZZ phase at $x=0.4$ [panels (a–c)] and the dZZ phase at $x=0.25$ [panels (d–f)] on an $8\times2\times12\times12$ cluster. 
We find $S_c^{zz}\simeq0.08$ for the g-ZZ phase and $S_c^{dzz}\simeq0.125$ for the dZZ phase. 
Panels (a,b) and (d,e) correspond to values of $S$ slightly below ($S\lesssim S_c$) and above ($S\gtrsim S_c$) these thresholds, while remaining in a non-condensed mean-field regime. 
For $S\lesssim S_c$, the spectra are gapped and exhibit precursor modes at the expected ordering wavevectors, namely the $M$ point for the g-ZZ phase and $M/2$ for the dZZ phase. 

Panels (c) and (f) correspond to the physical case $S=1/2$, which lies well above $S_c$ and thus deep in the ordered regime. 
To avoid saturation by the Bragg peak, only $0.1\%$ of the total spectral weight is shown, revealing the residual spectral weight associated with quantum fluctuations that persists even in these magnetically ordered states.

It is known that SBMFT is unable to describe  actual magnons since the gauge fluctuations leading to two-spinon bound states are neglected. 
Nevertheless the SBMFT $S({\bf q},\omega)$ predicts, in the ordered phases, a well-defined two-spinon dispersion which 
becomes singular in the thermodynamic limit \cite{Mezio_2012} and is consistent with the actual magnon dispersions obtained through series expansions \cite{Mezio_2011}. More refined SBMFT approaches incorporating gaussian fluctuations around the saddle-point do recover actual magnon \cite{Ghioldi_2018,Ghioldi_2022} as poles of the Random Phase Approximation (RPA) propagator. On the other hand, the non-singular SBMFT two-spinon continuum in $S({\bf q}, \omega)$
associated with unphysical fluctuations of the boson density constraint \cite{Mezio_2011} is removed at the RPA level\cite{Ghioldi_2022}.

Thus, the high intensity branches of Fig. \ref{fig07} obtained with the gSBMFT for $S>S_c$ can be interpreted as precursors of the actual magnons  (expected as two-spinon bound states in a theory beyond mean-field). Whether a two-spinon continuum coexists with the magnons in the exact 
excitation spectra of the ZZ and dZZ phases is unknown.

The dynamical signatures of the two phases remain clearly distinct. 
In particular, the dZZ phase exhibits a stronger spectral weight in the extended momentum region associated with the underlying triangular lattice, compared to the ZZ phase.
 

The magnetic structure of BaCo$_2$(AsO$_4$)$_2$ has only recently been clarified through polarized neutron-scattering experiments~\cite{regnault_polarized-neutron_2018}, which revealed a nearly commensurate double-zigzag order with wavevector $(0.27,0,-1.31)$. 
For comparison, the ideal commensurate dZZ state corresponds to the wavevector $(0.25,0,-1.31)$. 
Such an ordered state is expected to display a Goldstone mode near the $M/2$ point of the Brillouin zone, similar to the low-energy features observed in our Fig.~\ref{fig07}(f). In contrast, the INS intensity:
\begin{equation}
I(\mathbf{q},\omega) \propto 
\sum_{\alpha,\beta} 
\left( \delta_{\alpha\beta} - \frac{q_\alpha q_\beta}{|\mathbf{q}|^2} \right) 
S^{\alpha\beta}(\mathbf{q},\omega),
\label{eq:intensity}
\end{equation}
measured experimentally reveals a clear gap $\delta\simeq1.45$~meV at the $\Gamma$ point. This is also in contrast to our $S({\bf q}, \omega)$ of Fig. \ref{fig07}
where a minimum occurs at $M/2$ in the dZZ phase, as expected.

Recent classical Monte Carlo simulations of an extended $J$–$K$–$\Gamma$–$\Gamma'$ Kitaev–Heisenberg model provide a possible resolution of the apparent discrepancy between the measured static magnetic order and the excitation spectra. 
These calculations find an effective dZZ ground state with a slightly incommensurate wavevector $(0.27,0,-1.31)$, corresponding to a defective $uudd$ spin arrangement consistent with the experimental observations. 
Moreover, the simulated dynamical structure factor $S({\bf q},\omega)$ reproduces remarkably well the magnetic excitation spectra measured in INS experiments. 
The same model also accounts for the experimentally observed $1/3$ magnetization plateau associated with a $uud$ spin configuration.
However, it remains an open theoretical issue to understand the origin of the discrepancy between the defective dZZ order and the gapped mimimum of the excitation spectra observed at the $\Gamma$-point both in Monte Carlo simulations and experiments. On the other hand, linear spin wave theory is inappropriate for describing 
such defective sate.

\section{Conclusions and Outlook}
\label{sec:conclusion}

Motivated by the intriguing magnetic properties of BaCo$_2$(AsO$_4$)$_2$ (BCAO), we investigate the ferro–antiferromagnetic $J_1$–$J_3$ Heisenberg model on the honeycomb lattice. 
To properly address the frustrated ferromagnetism intrinsic to this model, we develop a generalized Schwinger-boson mean-field theory (gSBMFT) capable of treating FM and AFM interactions on an equal footing. 
Within this framework, the set of mean-field parameters is enlarged to include both triplet pairing and singlet hopping channels, in addition to the conventional singlet pairing and triplet hopping terms present in the standard SBMFT. 
This extension enables a unified description of competing exchange processes and allows us to determine the complete phase diagram of the $J_1$–$J_3$ Heisenberg model, together with its dynamical magnetic properties directly relevant to experimental observations on BCAO.

Classically, the model displays FM and ZZ phases (see Fig. \ref{fig05} (a)) at small and large
$x$, respectively, with an intermediate spiral phase appearing in a narrow
parameter region between them. 
In contrast, our phase diagram exhibits an intermediate dZZ phase,
consistent with our ED calculations on small clusters and previous
DMRG studies on finite cylinders (with up to $16^2$ sites) \cite{jiang_quantum_2023}. 
Since our approach can reach, in practical terms, the thermodynamic limit
(cluster sizes up to $48^2$), it provides strong evidence for the existence
of a dZZ phase (see Fig. \ref{fig05} (b))  in the $J_1$–$J_3$ model. 
The fact that this dZZ order differs from the classical intermediate spiral
phase highlights the prominent role of quantum fluctuations and points to
an underlying order-by-disorder mechanism that stabilizes the dZZ phase. 
We also note that the triplet pairing channels included in our approach are
crucial for stabilizing the dZZ phase, as they allow a consistent treatment
of ferromagnetic correlations within SBMFT at $T=0$ without violating the boson number constraint treated on average on arbitrarily large clusters~\cite{feldner_ferromagnetic_2011}.

We compute the magnetic excitation spectra of the model in order to make contact with neutron-scattering experiments. 
Magnetic order emerges above the critical spin values $S_c^{zz}\simeq0.08$ and $S_c^{dzz}\simeq0.125$ for the ZZ and dZZ phases, respectively. 
In this regime the spectra display minima at the corresponding ordering wavevectors $M$ and $M/2$, consistent with the Goldstone theorem. 
For $S<S_c$, however, the system remains non-condensed and a gapped quantum spin-liquid regime appears, accompanied by a redistribution of spectral weight between the first and second Brillouin zones, as illustrated in Fig.~\ref{fig07}. 
At the physical value $S=1/2$, the magnetic order is strongly developed and the spectra are dominated by Bragg peaks, although residual quantum fluctuations remain visible once the dominant weight is suppressed.

In contrast, inelastic neutron-scattering experiments on BCAO reveal a gapped magnon dispersion with a flat-band minimum near the $\Gamma$ point, reminiscent of a two-dimensional ferromagnet. 
This behavior is difficult to reconcile with elastic scattering experiments reporting incommensurate dZZ order at low temperature. 
The absence of the expected Goldstone mode close to the dZZ ordering wavevector therefore points to an apparent inconsistency between the observed static order and the excitation spectrum.

One possible resolution involves the role of disorder: the dZZ pattern may not be perfectly formed but instead contain defects, as recently suggested~\cite{devillez2025bonddependentinteractionsillorderedstate}. 
More generally, these seemingly contradictory observations may reflect the proximity of BCAO to a ferromagnetic instability. 
Indeed, experiments show that even a small magnetic field applied along the $b$ direction can drive the system into a ferromagnetic state, suggesting that BCAO behaves as a quasi-two-dimensional magnet close to a FM transition. 
Further clarification will require additional experimental studies, for instance using next-generation INS or resonant X-ray scattering techniques.

From a theoretical perspective, our work highlights the gSBMFT as a promising extension of SBMFT for investigating frustrated ferromagnets. 
It would be particularly interesting to study the effect of an external magnetic field and the sensitivity of the dZZ phase to it. 
Additional interactions relevant to BCAO, such as bond-dependent Kitaev terms and longer-range anisotropic couplings recently identified in Refs.~\cite{maksimov_baco_2aso_4_2_2025,devillez2025bonddependentinteractionsillorderedstate}, could in principle be incorporated as further refinements of the model. 
Finally, the effects of disorder and lattice distortions on the delicate balance between the ZZ, dZZ, and FM phases relevant to BCAO remain important directions for future work. These extensions would help clarify the microscopic mechanisms stabilizing the dZZ phase in BCAO and related cobalt-based honeycomb magnets.

\section*{Acknowledgment}
The authors are grateful to the anonymous referees for their careful reading
and constructive comments, which prompted us to revisit several aspects of
the work and significantly improve the manuscript.
A.R. would like to thank Virginie Simonet and Armand Devillez for stimulating
discussions. A.R. also acknowledges support from the French National Research
Agency through ANR FlatMoi, Grant No.~ANR-21-CE30-0029. J.M. acknowledges
financial support from MICIN/FEDER, Uni\'on Europea, under Grant
No.~PID2022-139995NB-I00, and from the Mar\'ia de Maeztu Programme for Units
of Excellence in R\&D, Grant No.~CEX2023-001316-M.
\appendix

\section{Connection to standard SBMFT}
\label{app.a}

In this Appendix, we clarify how the generalized formalism introduced in this work connects to the standard SBMFT framework. 
The expressions derived below are written in their most general form and include all pairing and hopping channels considered within the gSBMFT. 
The standard SBMFT is recovered as a particular limit obtained by restricting to singlet pairing and hopping channels.

We begin by considering the pairing basis 
$ \hat{\bf p}_{ij} = (\hat{p}_{ij}^{\ua\ua},\hat{p}_{ij}^{\ua\da},\hat{p}_{ij}^{\da\ua},\hat{p}_{ij}^{\da\da})$ 
and an isotropic exchange coupling $\hat{J} = \text{diag}(J^{xx},J^{yy},J^{zz})$ with $J^{xx}=J^{yy}=J^{zz}=J$. 
In this case, the interaction term can be written as 
\begin{eqnarray}
	\hat {\bf S}_i  \hat{ J}  \hat{\bf S}_j  =  \frac{1}{4}\hat{\bf p}_{ij}^{\dagger} T_p \hat {\bf p}_{ij},
\end{eqnarray}
with 
\begin{eqnarray}
	T_p  &=& \begin{bmatrix} 1 & 0 & 0 & 0 \\ 0 & -1 & 2 & 0  \\ 0 & 2 & -1 & 0 \\ 0 & 0 & 0 & 1  \end{bmatrix}.
\end{eqnarray}

We now define the singlet pairing operator through the projection
\begin{eqnarray}
	\hat{O}_A &=& \frac{1}{\sqrt{6}} \begin{bmatrix} 0 & 0 & 0 & 0 \\ 0 & 1 & -1 & 0 \\ 0 & -1 & 1 & 0 \\ 0 & 0 & 0 & 0 \end{bmatrix},
\end{eqnarray}
from which one obtains
\begin{eqnarray}
	\hat{A}^\dagger \hat{A} = - \frac{1}{4} \hat{\bf p}_{ij}^{\dagger} \hat{O}_A^\dagger T_p \hat{O}_A \hat{\bf p}_{ij},
\end{eqnarray}
where $\hat{A} = \frac{1}{2} (b_{i\ua}b_{j\da} - b_{i\da}b_{j\ua} )$ is the usual singlet pairing operator.

Similarly, in the hopping channel, we introduce the basis 
$ \hat{\bf h}_{ij} = (\hat{h}_{ij}^{\ua\ua},\hat{h}_{ij}^{\ua\da},\hat{h}_{ij}^{\da\ua},\hat{h}_{ij}^{\da\da})$ 
and define
\begin{eqnarray}
	\hat{B}^\dagger \hat{B}  = \frac{1}{4} \hat{\bf h}_{ij}^{\dagger} \hat{O}_B^\dagger T_h \hat{O}_B \hat{\bf h}_{ij},
\end{eqnarray}
with
\begin{eqnarray}
	\hat{O}_B = \frac{1}{\sqrt{6}} \begin{bmatrix} 1 & 0 & 0 & 1 \\ 0 & 0 & 0 & 0 \\ 0 & 0 & 0 & 0 \\ 1 & 0 & 0 & 1 \end{bmatrix},~~~ 
	T_h =  \begin{bmatrix} 1 & 0 & 0 & 2 \\ 0 &-1 & 0 & 0 \\ 0 & 0 &-1 & 0 \\ 2 & 0 & 0 & 1 \end{bmatrix}.
\end{eqnarray}
Here, $\hat{B} = \frac{1}{2}( b_{i\ua}^\dagger b_{j\ua} + b_{i\da}^\dagger b_{j\da})$ denotes the usual boson hopping operator.

Restricting the theory to the SU(2)-invariant pairing and hopping channels, one recovers the standard SBMFT relation
\begin{eqnarray}
	\hat {\bf S}_i \cdot \hat{\bf S}_j  = : \hat{B}^\dagger \hat{B} : - \hat{A}^\dagger \hat{A}.
\end{eqnarray}

This establishes the connection between the generalized formulation and the conventional SBMFT, which is obtained as a specific limit of the present theory.

More generally, the choice of projection operators $\hat{O}_{p,h}$ allows one to construct a broad class of mean-field Ans\"atze within a unified framework, encompassing both standard and generalized SBMFT descriptions.

\section{Diagonalizing the SBMFT}
\label{app.b}
We start by expressing the mean-field hamiltonian as:
\begin{equation}
H=\sum_{\bf q} \Psi^\dagger M_{\bf q} \Psi_{\bf q} + H'
\end{equation}
where: $\Psi_{\bf q}=(b^u_{{\bf q},\uparrow} b^v_{{\bf q}, \uparrow} b^{u\dagger}_{-{\bf q},\downarrow} b^{v\dagger}_{-{\bf q}, \downarrow} )^T $ and $M_{\bf q}$ a $4 \times 4$ matrix. The hamiltonian can be diagonalized through the transformation
$\Psi_{\bf q} = P_{\bf q} \Gamma_{\bf q}$:
\begin{equation}
    H=\sum_{\bf q} \Gamma^\dagger_{\bf q} M_{\bf q} \Gamma_{\bf q} + H'
\end{equation}
where $\Gamma_{\bf q}=(\gamma^1_{{\bf q},\uparrow} \gamma^2_{{\bf q}, \uparrow} \gamma^{1\dagger}_{-{\bf q},\downarrow} \gamma^{2\dagger}_{-{\bf q}, \downarrow} )^T $ 

which satisfies the two conditions: 
\begin{eqnarray}
P^\dagger_{\bf q} M_{\bf q} P_{\bf q} = \omega_{\bf q} 
\nonumber \\
P^\dagger_{\bf q} \tau^z  P_{\bf q} =\tau^z
\end{eqnarray}
where $\omega_{\bf q} =\text{diag}(\omega_1, \omega_2, \omega_1,\omega_2)$ is a diagonal matrix containing the bosonic eigen-frequencies and $\tau^z=\text{diag}(1,1,-1,-1)$.
This is equivalent to the generalized eigenvalue problem: $\tau^z M_{\bf q} P_{\bf q} = P_{\bf q} \tau^z \omega_{\bf q}$. Thus,
the transformation matrix $T_{\bf q} $ that satisfies the above two conditions simultaneously 
contains the eigenvectors of $\sigma^z M_{\bf q}$ in its columns and $\tau^z \omega_{\bf q}$ contains the bosonic eigen-frequencies
in pairs $(\omega_\mu, -\omega_\mu)$ with $\omega_\mu >0 $. After the numerical diagonalization of $\tau^z M_{\bf q}$, 
the eigenvectors, $\omega_\rho$, should be para-normalized so that they satisfy: 
$\omega^\dagger_\rho \tau^z \omega_\rho'=[\tau^z]_{\rho\rho'} $.

\section{Spin correlation calculations}
\label{app.c}
We briefly clarify the procedure used to compute spin--spin correlation functions within Schwinger-boson mean-field theory (SBMFT). It has recently been emphasized, in particular in the context of the Kitaev model~\cite{sasamoto2025schwingerbosontheorys1}, that the prescription adopted to evaluate
$
\langle {\hat {\bf S}}_i \cdot {\hat {\bf S}}_j \rangle
$
may significantly influence the physical interpretation of the results. Two approaches are commonly considered.\\
The first consists of an explicit evaluation of the four-boson operators in the Bogoliubov vacuum,
\begin{eqnarray}
\left\langle
\hat b_{i}^\dagger \sigma^\alpha \hat b_{i}\,
\hat b_{j}^\dagger \sigma^{\beta} \hat b_{j}
\right\rangle,
\end{eqnarray}
using the mean-field ground state without quasiparticle excitations, as derived for instance in Ref.~\cite{halimeh_spin_2016}.
The second approach relies on re-expressing the spin bilinears in terms of the same pairing and hopping operators employed in the mean-field decoupling of the Hamiltonian, schematically of the form
\begin{eqnarray}
\left\langle :\!\hat h^\dagger \hat h\!:\,-\hat p^\dagger \hat p \right\rangle .
\end{eqnarray}
Depending on the chosen prescription, certain contributions to the correlations may be partially or entirely neglected.\\
In the generalized framework (gSBMFT), the situation is more subtle. Nearest-neighbor and further-neighbor bonds are described by distinct sets of pairing and hopping operators, and therefore the spin bilinear $\mathbf{S}_i \cdot \mathbf{S}_j$ cannot, in general, be uniquely re-expressed in terms of a single mean-field channel. As a consequence, the decoupling-based reconstruction of correlations becomes ambiguous for complex Ans\"atze.\\
By contrast, the explicit evaluation of the bosonic operators in the Bogoliubov vacuum provides a controlled and unambiguous procedure. All physical information is encoded in the self-consistent saddle-point solution, and the correlation functions are obtained directly from the quadratic mean-field Hamiltonian. While both prescriptions coincide in standard SU(2)-symmetric SBMFT, the full vacuum evaluation becomes essential in the generalized framework.\\
Accordingly, throughout this work all real-space correlations, as well as static and dynamical spin structure factors, are computed from the complete expectation value of the boson operators in the Bogoliubov vacuum representing the mean-field ground state.\\

Following the general formulation of the spin structure factors \cite{halimeh2016}, we can derive the most general expression for our present four-boson generalized SBMFT. By definition the spin correlation between the component $\alpha$ on a reference site ${\bf r}_0+{\bf d}_u$ and $\beta$ on site ${\bf r}_j+{\bf d}_v$, where ${\bf d}_u$ is the sub-lattice vector, is given by
\begin{eqnarray}
\langle S_{{\bf r}_0+{\bf d}_u}^\alpha  S_{{\bf r}_0+{\bf d}_v}^\beta \rangle &=& \frac{1}{N} \sum_{i} \langle S_{{\bf r}_0+{\bf d}_u}^\alpha  S_{{\bf r}_0+{\bf d}_u+{\bf r}_j+{\bf d}_v}^\beta \rangle \\ &=& \frac{1}{N} \sum_{i} M_{\sigma_0 \sigma_1 \sigma_2 \sigma_3}^{\alpha \beta} \langle b_{i,\sigma_0}^{u+}b_{i,\sigma_1}^{u} b_{j,\sigma_2}^{v+}b_{j,\sigma_3}^{v} \rangle \nonumber
\end{eqnarray}
where translational invariance has been used and with $ M_{\sigma_0 \sigma_1 \sigma_2 \sigma_3}^{\alpha \beta} = \sigma_{\sigma_0 \sigma_1}^\alpha \otimes \sigma_{\sigma_2 \sigma_3}^\beta$ a $4 \times 4$ matrix defined from the Pauli matrices. Note that implicit summation is used for repeated spin indices $\sigma_i$. 

Performing the Fourier transform on the bosonic operators 
\begin{equation}
b^u_{i, \sigma}={1 \over \sqrt{N}} \sum_{{\bf q}} e^{i {\bf q} \cdot ( {\bf r}_i + {\bf d}_u ) } b^u_{{\bf q},\sigma},
\label{eq:ft}
\end{equation}
one can recast the correlation as 
\begin{widetext}
\begin{eqnarray}
\langle S_{{\bf r}_0+{\bf d}_u}^\alpha  S_{{\bf r}_j+{\bf d}_v}^\beta \rangle &=& \frac{1}{N^2}   M_{\sigma_0 \sigma_1 \sigma_2 \sigma_3}^{\alpha \beta}\sum_{{\bf q}_0,{\bf q}_1,{\bf q}_2} e^{i({\bf q}_0 - {\bf q}_1) \cdot ({\bf r}_j + {\bf d}_v - {\bf d}_u)}  \langle b_{{\bf q}_0,\sigma_0}^{u+}b_{{\bf q}_1,\sigma_1}^{u} b_{{\bf q}_2,\sigma_2}^{v+}b_{{\bf q}_3,\sigma_3}^{v} \rangle \nonumber \\
 &=& \frac{1}{N^2}   M_{\sigma_0 \sigma_1 \sigma_2 \sigma_3}^{\alpha \beta}\sum_{{\bf q}_0,{\bf q}_1,{\bf q}_2} e^{i({\bf q}_0 - {\bf q}_1) \cdot ({\bf r}_j + {\bf d}_v - {\bf d}_u)} \nonumber \\ &\times&  \sum_{\{l\}} \sum_{\{\sigma_l\}} 
 P_{(u,\sigma_0,1),(r,\sigma_r,0)}^{(-{\bf q}_0)} 
 P_{(u,\sigma_1,0),(s,\sigma_s,0)}^{({\bf q}_1)}
 P_{(v,\sigma_2,1),(m,\sigma_m,1)}^{(-{\bf q}_2)}
 P_{(v,\sigma_3,0),(n,\sigma_n,1)}^{({\bf q}_3)}  \langle \gamma_{-{\bf q}_0,\sigma_r}^{r}\gamma_{{\bf q}_1,\sigma_s}^{s} \gamma_{{\bf q}_2,\sigma_m}^{m+}\gamma_{-{\bf q}_3,\sigma_n}^{n+} \rangle \nonumber \\
 &+& \frac{1}{N^2}   M_{\sigma_0 \sigma_1 \sigma_2 \sigma_3}^{\alpha \beta}\sum_{{\bf q}_0,{\bf q}_1,{\bf q}_2} e^{i({\bf q}_0 - {\bf q}_1) \cdot ({\bf r}_j + {\bf d}_v - {\bf d}_u)} \nonumber \\ &\times&  \sum_{\{l\}} \sum_{\{\sigma_l\}} 
 P_{(u,\sigma_0,1),(r,\sigma_r,0)}^{(-{\bf q}_0)} 
 P_{(u,\sigma_1,0),(s,\sigma_s,1)}^{({\bf q}_1)}
 P_{(v,\sigma_2,1),(m,\sigma_m,0)}^{(-{\bf q}_2)}
 P_{(v,\sigma_3,0),(n,\sigma_n,1)}^{({\bf q}_3)}  \langle \gamma_{-q_0,\sigma_r}^{r}\gamma_{-q_1,\sigma_s}^{s+} \gamma_{-q_2,\sigma_m}^{m}\gamma_{-q_3,\sigma_n}^{n+} \rangle \nonumber
\end{eqnarray}
\end{widetext}
where ${\bf q}_3 = {\bf q}_2 +{\bf q}_0 -{\bf q}_1$, $\{l\}=(r,s,m,n)$, $\{ \sigma_l \}=(\sigma_r,\sigma_s,\sigma_m,\sigma_n)$, and where we have reexpressed the operators in terms of the eigenmodes:
\begin{eqnarray}
    \begin{bmatrix}
        b_{{\bf q},\sigma}^{u} \\ b_{-{\bf q},\sigma}^{u+}
    \end{bmatrix}
    = P^{({\bf q})}
    \begin{bmatrix}
        \gamma_{{\bf q},\sigma_r}^{r} \\ \gamma_{-{\bf q},\sigma_r}^{r+}
    \end{bmatrix}.
\end{eqnarray}
In this notation, an element of the eigen-matrix is given by $P_{(u,\sigma_u,a),(v,\sigma_v,b)}^{({\bf q})}$ with $a$ and $b$ = 0 or 1 depending on whether we consider annihilation operators ordered in the first $2 n_u$ elements, or creation ones in the last $2 n_u$.

From the above equations, only three contractions give non-zero expectation values:
\begin{eqnarray}
    C_1:~&&{\bf q}_2 = -{\bf q}_0,~~r=m,~~\sigma_r=\sigma_m,  \\
         &&{\bf q}_3 = -{\bf q}_1,~~s=n,~~\sigma_s=\sigma_n, \nonumber \\
    C_2:~&&{\bf q}_2 =  {\bf q}_1,~~s=m,~~\sigma_s=\sigma_m,  \\
         &&{\bf q}_3 =  {\bf q}_0,~~r=n,~~\sigma_r=\sigma_n, \nonumber \\
    C_3:~&&{\bf q}_0 =  {\bf q}_1,~~s=m,~~\sigma_s=\sigma_m,  \\
         &&{\bf q}_2 =  {\bf q}_3,~~r=n,~~\sigma_r=\sigma_n.\nonumber
\end{eqnarray}
Finally, noting that $P_{(u,\sigma_u,a),(v,\sigma_v,b)}^{(-{\bf q})} = {\bar P}_{(u,\sigma_u,{\bar a}),(v,\sigma_v,{\bar b})}^{({\bf q})}$, where we have defined ${\bar a} = (1,0)$ if $a=(0,1)$. we arrive to the final general expression 
\begin{widetext}
\begin{eqnarray}
\langle S_{r_0+\delta_u}^\alpha  S_{r_j+\delta_v}^\beta \rangle &=&\frac{1}{N^2}   M_{\sigma_0 \sigma_1 \sigma_2 \sigma_3}^{\alpha \beta}\sum_{{\bf q}_0,{\bf q}_1} e^{i({\bf q}_0 - {\bf q}_1) \cdot ({\bf r}_j + {\bf d}_v - {\bf d}_u)}  \nonumber \\
&\times& \sum_{r,s} \sum_{\sigma_r,\sigma_s} 
{\bar P}_{(u,\sigma_0,0),(r,\sigma_r,1)}^{({\bf q}_0)} 
 P_{(u,\sigma_1,0),(s,\sigma_s,0)}^{({\bf q}_1)}
 P_{(v,\sigma_2,1),(r,\sigma_r,1)}^{({\bf q}_0)}
 {\bar P}_{(v,\sigma_3,1),(s,\sigma_s,0)}^{({\bf q}_1)} \nonumber  \\
&+&\frac{1}{N^2}   M_{\sigma_0 \sigma_1 \sigma_2 \sigma_3}^{\alpha \beta}\sum_{{\bf q}_0,{\bf q}_1} e^{i({\bf q}_0 - {\bf q}_1) \cdot ({\bf r}_j + {\bf d}_v - {\bf d}_u)}  \nonumber \\
&\times& \sum_{r,s} \sum_{\sigma_r,\sigma_s} 
{\bar P}_{(u,\sigma_0,0),(r,\sigma_r,1)}^{({\bf q}_0)} 
 P_{(u,\sigma_1,0),(s,\sigma_s,0)}^{({\bf q}_1)}
 {\bar P}_{(v,\sigma_2,0),(s,\sigma_s,0)}^{({\bf q}_1)}
 P_{(v,\sigma_3,0),(r,\sigma_r,1)}^{({\bf q}_0)} \nonumber  \\
&+&\frac{1}{N^2}   M_{\sigma_0 \sigma_1 \sigma_2 \sigma_3}^{\alpha \beta}\sum_{q_0,q_1} \sum_{r,s} \sum_{\sigma_r,\sigma_s} 
{\bar P}_{(u,\sigma_0,0),(r,\sigma_r,1)}^{({\bf q}_0)} 
 P_{(u,\sigma_1,0),(r,\sigma_r,1)}^{({\bf q}_0)}
 {\bar P}_{(v,\sigma_2,0),(s,\sigma_s,1)}^{({\bf q}_1)}
 P_{(v,\sigma_3,0),(r,\sigma_r,1)}^{({\bf q}_1)} \nonumber
\end{eqnarray}    
\end{widetext}

When the system has no condensate (is gapped), the contraction $C_3$ is zero since it corresponds to onsite expectation values, namely the spin value at site ${\bf r}_i + {\bf d}_u$. It has been displayed for completeness of the general formulation. 

\begin{figure*}[t!]
  \includegraphics[clip, width=\textwidth]{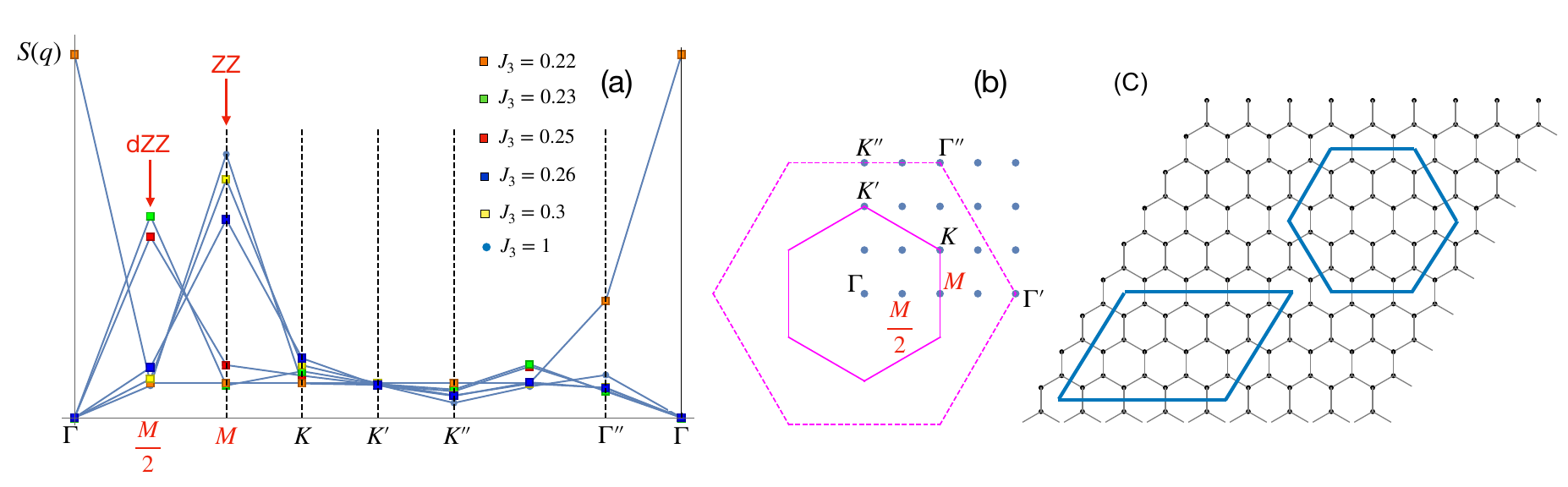}
  \caption{Magnetic orders in the $J_1-J_3$ model on a honeycomb lattice from ED. In (a) the ED static spin structure factor, $S({\bf q})$, on a $8 \times 3$ cluster at increasing $J_3$ is plotted along the path: $\Gamma-M-K-K'-K''-\Gamma''$, picking several symmetry points displayed in (b). The relevant $M=(2 \pi/\sqrt3,0)$ and $M/2$ wavevectors corresponding to the ZZ and dZZ magnetic orders are marked in red for clarity. The vertical dashed black lines mark the symmetry points. The solid (dashed) pink lines denote the 1st BZ of the honeycomb (completed triangular) lattice (see also Fig. \ref{fig01}) and the blue circles in (b) are the allowed wavevectors for the $8 \times 3$ cluster with PBC shown in (c). (c) The two 24-site clusters with different shapes and PBC used in the paper: the $8 \times 3$ (lower left), and the hexagonal fully symmetric  (upper right).}
  \label{fig:sqed}
\end{figure*}

\section{dZZ order from exact diagonalization}
\label{app.d}
In order to analyze the possible different magnetic orders in the $J_1$-$J_3$ Heisenberg model on the honeycomb lattice we have performed ED calculations on a 24-site ($8 \times 3$) cluster of Fig. \ref{fig:sqed} (c). 

The behavior of the static spin structure factor, $S({\bf q})$,  across the FM to ZZ transition is shown in Fig. \ref{fig:sqed}. 
While for $J_3<0.23$ a large peak at the $\Gamma$-point signaling the presence of FM order occurs, for $J_3>0.26$ a peak at the $M$-point associated with ZZ-order dominates $S({\bf q})$. This is consistent with the FM state expected for $J_3 \rightarrow 0$ 
and the ZZ order at large $J_3$. Strikingly, in the intermediate region $J_3 \approx 0.23-0.25$ a peak at the $M/2$-point corresponding to dZZ order emerges, dominating $S({\bf q})$.

Our results are consistent with previous DMRG calculations on large cylinders \cite{jiang_quantum_2023} with a similar cluster structure which have also found a small intermediate region where the dZZ state is stable. This also supports our SBMFT analysis in the main text. On the other hand, $S({\bf q})$ on the fully symmetric hexagonal 24-site cluster shown in Fig. \ref{fig:sqed} (c) does not add useful information on the dZZ order since it misses the $M/2$-point.

\bibliography{library}

\end{document}